\documentclass[11pt] {article}
\usepackage{float}
\usepackage{caption}
\usepackage{mathrsfs}
\usepackage{hyperref}
\usepackage{fancyhdr}
\fancypagestyle{firstpage}{
  \fancyhf{}
  
  \fancyhead[C]{\fontsize{9pt}{11pt}\selectfont
    This is the Author Accepted Manuscript (AAM) of the article published in \textit{Engineering Computations} (Emerald Publishing), 2025.\\
    The Version of Record is available at: \href{https://doi.org/10.1108/EC-06-2025-0583}{https://doi.org/10.1108/EC-06-2025-0583}
  }
}
\usepackage{graphicx} 
\usepackage{epstopdf}
\usepackage{graphics}
\usepackage{float}
\usepackage{amsmath}
\usepackage{graphicx}
\usepackage{multirow}
\usepackage{subcaption}
\usepackage{amsmath}
\usepackage{eufrak}

\DeclareMathAlphabet{\mathpzc}{OT1}{pzc}{m}{it}

\newtheorem{theorem}{Theorem}
\newtheorem{proof}{Proof}
\newcommand{\be}{\begin{equation}}
\newcommand{\ee}{\end{equation}}
\newcommand{\bea}{\begin{eqnarray}}
\newcommand{\eea}{\end{eqnarray}}

\usepackage{amsmath}
\usepackage{natbib}
\setcitestyle{authoryear,open={(},close={)}}
\usepackage{nomencl}
\usepackage{ifthen}

\renewcommand{\nomgroup}[1]{%
\ifthenelse{\equal{#1}{C}}{\item[\textbf{Abbreviations}]}{%
\ifthenelse{\equal{#1}{V}}{\item[\textbf{Plant parameters}]}{%
\ifthenelse{\equal{#1}{S}}{\item[\textbf{Plant and control parameters}]}{}}}
}
\makenomenclature
\begin{document}
\thispagestyle{firstpage}

\parindent=0pt
\parskip=.1cm
\begin{center}
\Large \bf 
Command-Filter-Based Trajectory-Tracking Control of Quadrotor subject to Internal and External Disturbances
\end{center}

\begin{center}
\vspace{8pt}
{\large Mustafa Mohammed Mustafa\textsuperscript{1,2}\par}

\vspace{4pt}

\textsuperscript{1}\,Department of Computer Science and Engineering, University of Kurdistan Hewlêr, Erbil, Iraq\par
\textsuperscript{2}\,Department of Electrical Engineering, Salahaddin University–Erbil, Erbil, Iraq\par
\end{center}

{ \Large \bf Abstract} \newline 
{ \Large \bf Purpose} \newline
We propose a command-filter backstepping controller that integrates a disturbance observer and an HGO to handle unknown internal and external disturbances acting on a quadrotor.
 \newline
{\Large \bf Design/methodology/approach} \newline
To build the controller, we first define tracking errors between the measured and desired quadrotor outputs. These errors let us rewrite the system in a new set of state variables. Using that transformed model, we apply the Lyapunov theory and derive a backstepping control law. To avoid taking repeated time-derivatives of states and virtual controls, we insert a first-order command filter. Since the controller also needs disturbance estimates, we add a nonlinear DO. Finally, we replace every state that appears in the controller or observer with its estimate from an HGO.
\newline
{\Large \bf Findings} \newline
The main result is a control law that lets the quadrotor follow its path even when both internal and external disturbances act on it. Every sub-model is allowed its own type of disturbance, so the design stays realistic. We introduce a new state transformation and, with Lyapunov arguments, build a backstepping controller that includes a first-order filter; the filter keeps the design from suffering the usual “explosion of complexity.” An HGO then reconstructs the unmeasured states and their rates, yielding an output-feedback implementation. In parallel, the nonlinear DO attenuates constant and nonlinear disturbances and band-limited white noise.
\newline
{\Large \bf Practical implications}  \newline
The method reduces reliance on high-precision sensors and mitigates the impact of wind, model error, and rotor noise during flight.
 \newline
{\Large \bf Originality/value} \newline
Previous studies typically address either disturbance rejection or partial sensing, rarely both. Our design brings the filter, DO, and HGO together, so it tackles disturbances, limited sensors, and the well-known complexity spike in backstepping all at once.
\newline
 {\Large \bf Keywords} \newline
Output-feedback control, Backstepping control, Nonlinear DO, HGO, Quadrotor
\section{Introduction}\label{sec:i11ntro}

Unmanned aerial vehicles (UAVs) attract considerable research interest because they can hover, glide, and navigate without an onboard pilot. Current uses range from search-and-rescue and site inspection to payload delivery, wildfire monitoring, and defense \citep{quan2017introduction}. Platforms are typically divided into fixed-wing aircraft, which require runways, and rotary-wing craft that take off vertically. This study focuses on the quadrotor, a four-rotor vehicle that is underactuated, possessing only four control inputs for six motion variables.

This research focuses on the quadrotor's control design, a complex and coupled system featuring four rotors equidistant from its center of gravity. This configuration consists of two perpendicular arms. This underactuated model has four inputs and six outputs, encompassing Cartesian position and attitude angles \citep{mahony2012multirotor}. Consequently, effective quadrotor control uses a two-loop structure: an outer position loop and an inner attitude loop. Position control not only manages quadrotor position but also generates desired attitude angles for attitude control. Hence, position control acts as the outer loop, and attitude control as the inner loop \citep{gajbhiye2022geometric,xia2017robust}.

Linear controllers for quadrotor control, such as proportional-integral-derivative (PID), proportional-derivative (PD), and linear quadratic regulator (LQR) control, use outputs and desired references to compute errors and develop linear control algorithms \citep{bouabdallah2004pid,tayebi2006attitude}. These controllers are designed based on linearized mathematical models that assume initial conditions near the origin. To address limitations of linear control, nonlinear control methods have been implemented for quadrotor control, demonstrating performance improvements.

Nonlinear control approaches and Lyapunov stability analysis are effective for advanced control systems under conditions of uncertainty. They can manage uncertainties while maintaining system stability \citep{mustafa2020dedicated, SU2020}. Sliding mode control (SMC) is a robust nonlinear control method for stabilizing and tracking the trajectory of quadrotor UAVs. \cite{zhao2015nonlinear} designed an attitude control using standard SMC, achieving significant performance through a two-step process: constructing a sliding manifold and enforcing attitude angles to slide on it. \citep{reinoso2016trajectory} developed an SMC algorithm for point-to-point trajectory tracking by addressing both position and attitude subsystems. Other notable works include \citep{perozzi2018trajectory,ai2019fixed,xu2006sliding}. However, these techniques often suffer from chattering, which can damage actuators. Additionally, ensuring robustness requires a known upper bound on external disturbances.

Researchers have reduced SMC chattering with high-order SMC and super-twisting algorithms as presented in \cite{chen2021attitude,utkin2020conventional}. \citep{bouabdallah2005backstepping} introduced the backstepping control technique for achieving trajectory tracking, involving the design of an auxiliary variable based on measurable and desired state errors, followed by Lyapunov criteria to ensure stable system dynamics. A finite-time controller is designed to handle bounded uncertainties and disturbances, while a barrier Lyapunov function enforces vertical descent acceleration constraints within a safe bound \citep{khadhraoui2023barrier}. \citep{wang2018trajectory} tackled input saturation issues and achieved trajectory tracking through backstepping control, noting complexities arising from repeated derivatives of control inputs \citep{CHEN2014436}. However, these methods typically operate under ideal conditions, without addressing external and internal disturbances encountered during flight missions. Furthermore, they often require empirical tuning of control gains, posing challenges in dynamic and uncertain environments.

To address disturbances and uncertainties and reduce empirical tuning, adaptive control techniques have been integrated with backstepping and SMC methods \citep{koksal2020backstepping,li2021command, zhu2025event}. These techniques use state-dependent control gain criteria derived from a Lyapunov candidate function. Additionally, neural networks are incorporated to minimize chattering in standard SMC. Another popular method is model-reference adaptive control (MRAC). \citep{6978900} proposed MRAC algorithms for quadrotor trajectory tracking, designing a reference model for desired trajectories, and using the error between real and reference outputs to satisfy Lyapunov criteria and finalize adaptive control laws. Other robust controllers that address disturbances are presented in \cite{9642982, LOPEZSANCHEZ2021243}, but they use a DO for mitigating external and internal disturbances.

Wind gusts, actuator/rotor noise, and modeling uncertainty are inherent in UAV flight control \citep{li2025finite}. To estimate and reject these effects, \citep{chen2000nonlinear} introduced the nonlinearDO and its integration with a nominal controller in disturbance observer-based control (DOBC), where the observer augments the controller and the nominal law remains unchanged when disturbances vanish. For payload-carrying missions, wind-gust estimation criteria were developed in \citep{wang2016trajectory}. Disturbances can be separated into internal (parametric) and external (environmental) components, and DO-based methods have been used to estimate and attenuate both \citep{antonelli2017adaptive}. For aggressive flight, incremental nonlinear dynamic inversion with differential flatness achieves accurate tracking by decoupling channels through inverse-dynamics compensation \citep{chen2019tracking}, and a recent survey consolidates advances in Incremental Nonlinear Dynamic
Inversion (INDI) while emphasizing robustness to modeling errors and sensor noise \citep{steinert2025fundamentals}. Complementary results show that barrier-Lyapunov backstepping combined with command filtering and per-axis nonlinear DOs can enforce prescribed attitude-error bounds under physical and virtual input saturations, with experimental validation \citep{jouffroy2025robust}.

The aforementioned DOBC techniques relied heavily on accurate state variable and rate information, requiring efficient onboard sensors, which may not always be available \citep{siddiqui2024model}. To reduce sensor dependency, output feedback control (OFC) using an HGO combined with SMC techniques and DO was proposed by \citep{ahmed2022adaptive}. This approach enabled quadrotor attitude control with adaptive laws based on estimated state variables, departing from traditional methods.  \citep{shao2018rise} extended HGO to EHGO for state estimation, integrating it with backstepping control, albeit limited to integral-chain systems. Despite their effectiveness, both state estimation techniques exhibited peaking phenomena due to inherent high-gain settings before reaching steady state.

To handle these issues we propose a control scheme that rejects both external disturbances and internal model errors. Additionally, a robust state estimation criterion has been designed to enhance reliance on estimated states, reducing dependence on precise sensor measurements. Furthermore, the proposed control algorithm is designed to achieve precise trajectory tracking along all XYZ axes. The primary contributions of this research include:

\begin{enumerate}
    \item One DO across loops. A single nonlinear DO estimates and rejects multiple disturbance types (exogenous, inverted-ramp, chirp, band-limited) in both attitude and position channels.
    
    \item Consistent observer use. An HGO provides state estimates that are used both in the DO and in the control law, which lowers dependence on accurate sensors.
    
    \item End-to-end underactuated design. Command-filtered backstepping with motor mixing maps virtual position controls to desired attitudes and then to the four actuator inputs, avoiding repeated derivatives while retaining the Lyapunov steps.
\end{enumerate}

\section{Mathematical model and preliminaries }\label{sec:problem_description }
\subsection{Mathematical model}
According to \cite{zuo2010trajectory}, the quadrotor dynamics are:
\begin{equation} \label{eq:Quadrotor_Dynamical_Model}
\begin{split}
\text{Position model: } & \begin{cases}
\ddot{x}  = \left( \cos \phi \sin \theta \cos \psi + \sin \phi \sin \psi  \right) {U_p}/{m} \\
\ddot{y}  = \left( \cos \phi \sin \theta \sin \psi - \sin \phi \cos \psi \right) {U_p}/{m} \\
\ddot{z}  = \left( \cos \phi \cos \theta \right) {U_p}/{m} - g 
\end{cases} \\
\text{Attitude model: } & \begin{cases}
\ddot{\phi}  = \dot{\theta} \dot{\psi}  \left( {I_y - I_z} \right) / {I_x}   + \dot{\theta} \Omega_r {I_r}/{I_x}  + {l}/{I_x} U_\phi;  \\
\ddot{\theta}  = \dot{\phi} \dot{\psi} \left(  {I_z - I_x}\right) /{I_y} - \dot{\phi} \Omega_r {I_r}/{I_y} + {l}/{I_y} U_\theta;  \\
\ddot{\psi}  = \dot{\phi} \dot{\theta} \left(  {I_x - I_y}\right) /{I_z}  +  {1}/{I_z} U_\psi
\end{cases}
\end{split}
\end{equation}
where $\phi$, $\theta$, and $\psi$ denote the roll, pitch, and yaw angles, respectively, constrained to $(-\pi/2, \pi/2)$. The variables $x$, $y$, and $z$ denote the quadrotor's position along the x, y, and z axes. Quadrotor parameters used in this study are presented in  Table \ref{tab:Quadrotorparamters_5}. Control variables are defined as follows \citep{bouabdallah2005backstepping, zuo2010trajectory}.
\begin{table}[H]
  \centering
  \caption{Quadrotor parameters. \\ \textit{Source: Author’s own work; parameters adapted from \cite{zuo2010trajectory}}}
  \label{tab:Quadrotorparamters_5}
  \begin{tabular}{ l l l l }
    \hline\hline
    Parameter & Symbol & Value & Unit \\ \hline
    Gravity & $g$ & $9.81$ & $\mathrm{m.s^{-2}}$ \\
    Mass & $m$ & $0.650$ & $\mathrm{kg}$ \\
    Arm length (CoM to rotor) & $l$ & $0.235$ & $\mathrm{m}$ \\
    Thrust coefficient & $b$ & $2.980 \times 10^{-6}$ & $\mathrm{N.s^{2}}$ \\
    Drag (torque) coefficient & $d$ & $7.5 \times 10^{-7}$ & $\mathrm{N.m.s^{2}}$ \\
    Rotor inertia & $J_r$ & $3.357 \times 10^{-5}$ & $\mathrm{kg.m^{2}}$ \\
    Airframe inertia (roll) & $I_x$ & $7.5 \times 10^{-3}$ & $\mathrm{kg.m^{2}}$ \\
    Airframe inertia (pitch) & $I_y$ & $7.5 \times 10^{-3}$ & $\mathrm{kg.m^{2}}$ \\
    Airframe inertia (yaw) & $I_z$ & $1.3 \times 10^{-3}$ & $\mathrm{kg.m^{2}}$ \\
    Linear drag (x-axis translation) & $A_x$ & $0.25$ & $\mathrm{kg.s^{-1}}$ \\
    Linear drag (y, pitch axis translation) & $A_y$ & $0.25$ & $\mathrm{kg.s^{-1}}$ \\
    Linear drag (z, vertical translation) & $A_z$ & $0.25$ & $\mathrm{kg.s^{-1}}$ \\
    Motor inertia & $I_M$ & $3.357 \times 10^{-5}$ & $\mathrm{kg.m^{2}}$ \\
    Residual angular speed & $\Omega_r$ & $-$ & $\mathrm{rad.s^{-1}}$ \\
    \hline
  \end{tabular}
\end{table}

\begin{equation} \label{eq:Control_inputs}
\begin{split}
U_p &  = b \left(  \omega_1^2 + \omega_2^2 + \omega_3^2 + \omega_4^2 \right) \\
U_\phi & = b \left( \omega_2^2 + \omega_4^2 \right) \\
U_\theta & = b \left( \omega_1^2 + \omega_3^2 \right) \\
U_\psi  &= d \left( \omega_1^2 - \omega_2^2 + \omega_3^2 - \omega_4^2 \right) 
\end{split}
\end{equation}

Define $\phi = x_1 $, $\dot{\phi} = x_2 $, $\theta = x_3 $, $\dot{\theta}=x_4 $, $\psi = x_5  $, $\dot{\psi} = x_6  $, $x = x_7  $, $\dot{x}=x_8 $, $y = x_9 $, $\dot{y}= x_{10} $, $z = x_{11}  $ and $\dot{z} = x_{12} $.  Next,  define $X = [
x_1 , x_2 , x_3 , x_4 , x_5 , x_6 ,x_7 , x_8 , x_9 , x_{10} , x_{11} , x_{12}
] $. Hence, a simplified state-space model can be obtained as follows:
\begin{equation} \label{eq:Attitude_State_space_model}
\dot{X}=   
\begin{bmatrix}
\dot{x}_1   \\ \dot{x}_2   \\ \dot{x}_3   \\ \dot{x}_4   \\ \dot{x}_5   \\ \dot{x}_6  \\ \dot{x}_7   \\ \dot{x}_8   \\ \dot{x}_9   \\ \dot{x}_{10}   \\ \dot{x}_{11}   \\ \dot{x}_{12}  
\end{bmatrix} 
= 
\begin{bmatrix}
x_2  \\ x_4   x_6  (I_y-I_z)/I_x + x_4   \Omega_r I_r/I_x + l/I_x U_\phi      \\
x_4  \\ x_2  x_6  (I_z-I_x)/I_y - x_2  \Omega_r I_r/I_y + l/I_y U_\theta       \\
x_6  \\ x_2  x_4  (I_x - I_y )/I_z+ 1/I_z  U_\psi    \\
x_8   \\  (\cos x_1 \sin x_3 \cos x_5 + \sin x_1 \sin x_5) U_p/m \\
x_{10}   \\  (\cos x_1 \sin x_3 \sin x_5 - \sin x_1 \cos x_5)U_p/m \\
x_{12}   \\ - g + \left( \cos x_1 \cos x_3 \right) {U_p}/{m} 
\end{bmatrix}
\end{equation}
where the auxiliary control inputs for the x-axis and y-axis are assumed as follows:
\begin{equation} \label{eq:Ux_Uy_conts}
\begin{split}
U_{x} & = \cos x_1 \sin x_3 \cos x_5 + \sin x_1 \sin x_5 \\
U_{y} & = \cos x_1 \sin x_3 \sin x_5 - \sin x_1 \cos x_5 
\end{split}
\end{equation}

\subsection{Lemmas and assumptions}
\textbf{Lemma 1:} According to the separation principle, output-feedback control achieves the same trajectories as state-feedback control if the state-estimation gains are sufficiently large. \citep{khalil2007high}. \newline
\textbf{Assumption 1:} Quadrotor position and attitude are measurable.
\section{Control development and stability analysis}
\subsection{Attitude Control}
The roll submodel of the attitude system in a quadrotor can be written in a class of nonlinear systems as follows:
\begin{equation}
\begin{split}
\dot{x}_1   =\ & x_2  \\
\dot{x}_2  =\ & \frac{1}{I_x} \left( \left( {I_y - I_z} \right)x_4  x_6   + {I_r} \Omega_r  x_4  \right)+ \frac{1}{I_x}l U_\phi  + d_\phi 
\end{split} \label{eq:Class_of_NL_sys}
\end{equation}
 where $d_\phi $ is an unknown disturbance.  First, a command filter for the desired roll angle and its rate is designed as follows:
 \begin{equation} \label{eq_comm_fill_roll}
 \begin{split}
 \dot{z}_1 &= -m_1 |z_1 - x_{\phi_r}|^{1/2} \text{sign} \left( z_1 - x_{\phi_r} \right) + z_2 \\
 \dot{z}_2 &= -m_2 \text{sign} \left(z_1 - x_{\phi_r} \right)
 \end{split}
 \end{equation}
 where $m_1>0$ and $m_2$ are design constants. $z_1 = x_1^c $ and $z_2 = x_2^c = \dot{x}_{\phi_r}$. Define the tracking error as
\begin{equation}
\xi_1  = x_1  - x_1^c 
\end{equation}
The time derivative is obtained as
\begin{equation}
\dot{\xi}_1  = \dot{x}_1  - x_2^c  = x_2   - x_2^c   \label{eq:z_1_dot_old}
\end{equation}
Auxiliary control is designed as
\begin{equation}
\nu_1  = -p_1 \xi_1  + \dot{x}_{\phi r}
\end{equation}
where $p_1 >0$ is a design constant. Next, the filter is constructed as
\begin{equation}
\begin{split}
\tau \dot{\varsigma_1}  + \varsigma_1  =\ &  \nu_1 ; \ \ \text{where} \ \ \varsigma_1 (0) =  \nu_1 (0)
\end{split}
\end{equation}
where $\tau$ and $\varsigma_1$  represent the time constant and filter, respectively. Next, an error between filter and auxiliary controller is given by   $e_1 = \varsigma_1  - \nu_1 $.  Thus, $\xi_2  $ is obtained as 
\begin{equation}
\xi_2   = x_2  - \varsigma_1  -  x_2^c  \label{eq:z_2_old}
\end{equation}
Simplified  (\ref{eq:z_1_dot_old}) is given by
\begin{equation}
\dot{\xi}_1  = - {p}_1 \xi_1  + \xi_2  + e_1 +  x_2^c   \label{eq:z_1_dot_modified_old}
\end{equation}
Time derivative of $\xi_2$ is derived as
\begin{equation}
\dot{\xi}_2  = \frac{1}{I_x} \left( \left( {I_y - I_z} \right)x_4  x_6   + {I_r} \Omega_r  x_4  \right) + \frac{1}{I_x}l U_\phi  + d_\phi   -  \dot{x}_2^c  - {\nu_1 }/{\tau} + {\varsigma_1 }/{\tau} \label{eq:z_2_dot_old}
\end{equation}
Next, the Lyapunov candidate function is chosen as
\begin{equation}
V_{\text{A}}  = 0.5( \xi_1^2  +  \xi_2^2  + e_1^2   +  \tilde{d}_{\phi}^2 ) \label{eq:Lyap_fn_old}
\end{equation}
where    $\tilde{d}_\phi  = \hat{d}_\phi  - d_\phi   $   and $\hat{d}_\phi $  is the estimation of disturbance obtained according to the DO proposed by \cite{chen2000nonlinear}, written as follows:
\begin{equation}
\begin{split}
\dot{\gamma}_\phi  =\ &  - l_\phi(x) g_{2\phi}     \gamma_\phi   - l_\phi(x) \big(g_{2\phi}    p_\phi(x) + h_\phi(x)+ g_{1\phi }     U_\phi  \big); \\
\hat{d}_\phi  =\ &  \gamma_\phi   + p_\phi(x)
\end{split} 
\end{equation}
where $p_\phi(x)$ is a nonlinear design function.  $l_\phi(x)$, $g_{1\phi}   $ and $g_{2\phi}   $ are obtained as follows \begin{equation}
\begin{split}
& l_\phi(x) = \frac{\partial p_\phi(x)}{\partial x}; \ \ h_\phi(x) = \begin{bmatrix}
x_2   & \frac{1}{I_x} \left( \left( {I_y - I_z} \right)x_4  x_6   + {I_r} \Omega_r  x_4  \right)
\end{bmatrix}^T \ \\
& g_{1\phi}   = \begin{bmatrix}
0 & \frac{1}{I_x}l
\end{bmatrix}^T; \ \ g_{2\phi}   = \begin{bmatrix}
0 & 1
\end{bmatrix}^T
\end{split}
\end{equation}
 Next,  derivative of (\ref{eq:Lyap_fn_old}) is derived as follows
\begin{equation} \label{eq:V_dot_attitude}
\dot{V}_{\text{A}}  = \xi_1  \dot{\xi}_1  + \xi_2  \dot{\xi}_2  + e_1 \dot{e}_1   + \tilde{d}_\phi  \dot{\tilde{d}}_\phi  
\end{equation}
where $\dot{\tilde{d}}_\phi   = \dot{\hat{d}}_\phi   - \dot{d}_\phi  $ can be obtained as follows:
\begin{equation}
\dot{\tilde{d}}_\phi   = - l_\phi (x)g_{2 \phi}   \tilde{d}_\phi  = - \lambda_\phi \tilde{d}_\phi  \label{eq:tilde_d_old }
\end{equation}
Substitute (\ref{eq:z_1_dot_modified_old}), (\ref{eq:z_2_dot_old}) and (\ref{eq:tilde_d_old })  into (\ref{eq:V_dot_attitude}) 
\begin{equation}
\begin{split}
\dot{V}_{\text{A}}  
= \ & -p_1 \xi_1^2  + \xi_1  \xi_2  + \xi_1  e_1 + \xi_1   x_2^c   + \xi_2  \big(\frac{1}{I_x} \left( \left( {I_y - I_z} \right)x_4  x_6   + {I_r} \Omega_r  x_4  \right) + \frac{1}{I_x}l  U_\phi \\
\ &   + d_\phi   -  \dot{x}_2^c   - {\nu_1 }/{\tau} + {\varsigma_1 }/{\tau} \big) - {e_1^2}/{\tau}    - F_{\text{A}}(\star) e_1    - \lambda_\phi \tilde{d}_\phi ^2  
\end{split} \label{eq:derivative_of_LF_old}
\end{equation}
where $F_{\text{A}}(\star)  = F_{\text{A}}(x_1,x_2,\xi_1,\xi_2,e_1) = \frac{\partial \nu_1 }{\partial x_1} \dot{x}_1  + \frac{\partial \nu_1}{\partial \xi_1} \dot{\xi}_1$ is a continuous function. By \cite{wang2005neural}, if the initial conditions of $F_{\text{A}}(\star)$ lie in a compact set, then it can be bounded by a constant $\bar{F}_{\text{A}}$.
The control law to achieve the stable dynamics is proposed as follows
\begin{equation} \label{eq:standard_BSC_old}
\begin{split}
U_\phi    = &  - \left( {\frac{1}{I_x}l} \right)^{-1} \big(\xi_1  + \frac{1}{I_x} \left( \left( {I_y - I_z} \right)x_4  x_6   + {I_r} \Omega_r  x_4  \right) -  \dot{x}_2^c   - ( {\nu}_1 - \varsigma_1  )/{\tau} + k_1  \xi_2  + \hat{d}_\phi    \big)
\end{split}
\end{equation}
where $k_1>0$. Substitute (\ref{eq:standard_BSC_old}) into (\ref{eq:derivative_of_LF_old}) and use Young's inequality 
\begin{equation}
\begin{split}
\dot{V}_{\text{A}}  
\leq \ &  -\left( p_1 - 1/2 \right) \xi_1^2   - \left( k_1 -1/2 \right) \xi_2^2   - (1/\tau - 1/2)e_1^2 - \left(\lambda_\phi - 1/2 \right) \tilde{d}_\phi^2     +  B_1/2
\end{split} \label{eq:V_dot_LF_Adapr}
\end{equation}
where $B_1 \geq \text{max} \{ 3  (x_2^c)^2     + \bar{F}_{\text{A}}^2   \}$.  Therefore, by choosing positive control gains and selecting $p_\phi(x)$ so that $l_\phi(x)g_{2\phi}> {1}/{2}$, we ensure $\dot{V}_{\text{A}}<0$, resulting in an asymptotically stable closed-loop system.

Next, $d_{\phi}   = 0$ is assumed for the design of state-estimation. Then, an HGO for state estimation is designed as follows \cite{khalil2007high}:
\begin{equation} \label{eq:HGO_roll}
\begin{split}
\dot{\hat{x}}_1 &= \hat{x}_2 + {\beta_{1}} \tilde{x}_1/\varepsilon\\ 
\dot{\hat{x}}_2 &= \hat{f}_1(\hat{x} ) + \frac{1}{I_x}l U_\phi + {\beta_{2}} \tilde{x}_1 /{\varepsilon}^2
\end{split}
\end{equation}
where $\tilde{x}_1 = y_{\phi} - \hat{x}_1$ and $y_{\phi}=x_1$, $\hat{f}_1(\hat{x})$ is nominal model of $f_1(x )=\frac{1}{I_x} \left( \left( {I_y - I_z} \right)x_4  x_6   + {I_r} \Omega_r  x_4  \right)$ and $\varepsilon>0$ is a constant. Furthermore, $\beta_{1}>0$ and  $\beta_{2}>0$ are design constants. 
Next, state-estimation errors are defined as follows
\begin{equation}
\chi_1 =\tilde{x}_1/{\varepsilon} ; \ \chi_2 = \tilde{x}_2
\end{equation}
Define $\chi=[\chi_1,\chi_2]^T$, thus $\varepsilon \dot{\chi} = A\chi + \varepsilon \delta $, where matrices    $A$ and $\delta   $ are written as follows:
\begin{equation}
A = \begin{bmatrix}
- \beta_{1} & 1;  - \beta_{2} & 0
\end{bmatrix};
\delta   = \begin{bmatrix}
0 & \tilde{f}
\end{bmatrix}^T
\end{equation}
where $\tilde{f}= f_1(x )-\hat{f}_1(\hat{x} )$ and $A$ is Hurwitz due to the characteristic polynomial of $\beta_{1}$ and $\beta_{2}$. Choose the Lyapunov candidate
\begin{equation} \label{eq:HGO_roll2}
V_{H} = \chi^TP\chi
\end{equation}
  where $P=[p_{11},p_{12};p_{21},p_{22}]$ obtained by solving $PA+A^TP=-I$ with constants such that $P=P^T>0$ and $p_{12}=p_{21}$. 
Time derivative of (\ref{eq:HGO_roll}) is obtained as follows
\begin{equation} \label{eq:HGO_LF_drivative}
\begin{split}
\dot{V}_H
&\leq -(1-0.5\varepsilon^2 )\chi_1^2 - \kappa_\phi \chi_2^2  +\tilde{f}^2
\end{split}
\end{equation}
where $\kappa_\phi  = (1- \varepsilon^2((1+\beta_{2})/\beta_{1}+ \beta_{1}))/(2\beta_{2}^2))$.
Therefore, asymptotic stability can be achieved by appropriate choice of $\beta_{1}>0$, $\beta_{2}>0$, and designing $\varepsilon \in (0,1]$ sufficiently small. The asymptotic stability is also investigated and validated in the following corollary.

The following theorem summarizes the design.

\begin{theorem}
For a class of nonlinear systems in   (\ref{eq:Class_of_NL_sys}), a backstepping dynamic surface-based control combined   and DO can be designed as follows:
\begin{equation}\label{eq:controller_theorem_updated_old}
\begin{split}
  U_\phi   =& \ - \left( {\frac{1}{I_x}l} \right)^{-1} \big(\xi_1  + \frac{1}{I_x} \left( \left( {I_y - I_z} \right) \hat{x}_4  \hat{x}_6   + {I_r} \Omega_r  \hat{x}_4  \right)   - \left(  {\nu}_1 - \varsigma_1-1 \right)/{\tau}   + {k}_1  \xi_2  + \hat{d}_\phi   \big) \\
  \dot{\gamma}_\phi  =& \   - l_\phi(\hat{x}) g_{2\phi}   \gamma_\phi   - l_\phi\big(\hat{x}) ( g_{2\phi}  p_\phi(\hat{x})  + h_\phi (\hat{x})  + g_{1\phi}   U_\phi \big); \ 
\hat{d}_\phi  =  \gamma_\phi   + p_\phi(\hat{x})   \\ 
 {\nu}_1    =& \ -  {p}_1 \xi_1 +  x_2^c; \   \dot{\varsigma}_1  =    \left(  {\nu}_1   - \varsigma_1   \right)/{\tau} ;     \\
 \dot{\hat{x}}_1  =& \  \hat{x}_2 + {\beta_{1}} \tilde{x}_1/{\varepsilon} ; \
\dot{\hat{x}}_2   = \hat{f}_1(\hat{x} ) + \frac{1}{I_x}l U_\phi  + {\beta_{2}}\tilde{x}_1/{\varepsilon} ^2
\end{split}
\end{equation}
\end{theorem}
\begin{proof}
 Define Lyapunov candidate function
 as follows:
\begin{equation}
V = V_A + V_H
\end{equation}
Obtain derivative and substitute (\ref{eq:V_dot_LF_Adapr}) and (\ref{eq:HGO_LF_drivative})  
\begin{equation} \label{eq:Theorem_proof}
\begin{split}
\dot{V} \leq \ & -\left( p_1 - 1/2 \right) \xi_1^2   - \left( k_1 -1/2 \right) \xi_2^2   - (1/\tau - 1/2)e_1^2 - \left(\lambda_\phi - 1/2 \right) \tilde{d}_\phi^2 \\
\ & -(1-0.5\varepsilon^2 )\chi_1^2 - \kappa_\phi \chi_2^2  +\tilde{f}^2
\end{split}
\end{equation}
Therefore, with the appropriate control gain design for the controller, DO and HGO achieve asymptotically closed-loop stability. This completes the proof.
\end{proof}

For pitch and yaw, a similar control development approach can be followed. 
\subsection{Position control}
The position state-space model is underactuated with one control input, $U_p$, and three outputs $x$, $y$, and $z$ as shown in (\ref{eq:Quadrotor_Dynamical_Model}). To develop a position controller, We define an auxiliary control  $U = [U_x, U_y, U_z]^T =  -gz + \frac{T}{m} Rz  $ \cite{zuo2010trajectory}.

 The command filter for the desired position along the x-axis and its rate is designed as follows:
 \begin{equation} \label{eq_comm_fill_pos_x}
 \begin{split}
 \dot{z}_7 &= -m_7 |z_7 - r_x|^{1/2} \text{sign} \left( z_7 - r_x \right) + z_8 \\
 \dot{z}_8 &= -m_8 \text{sign} \left(z_7 - r_x \right)
 \end{split}
 \end{equation}
 where $m_7>0$ and $m_8>0$ are design constants. $z_7 = x_7^c $ and $z_8 = x_8^c = \dot{r}_{x}$.  Next,  tracking error is defined as follows:
\begin{equation}
\xi_7 = x_7 - x_7^c 
\end{equation}
where $r_x $ represents the desired position along the x-axis. Derivative is obtained as follows
\begin{equation}
\dot{\xi}_7 = \dot{x}_7 - \dot{r}_x = x_8  - x_8^c  \label{eq:z_1_dot}
\end{equation}
Auxiliary control law is designed as 
\begin{equation}
\nu_4 = -p_4 \xi_7 + x_8^c 
\end{equation}
where $p_4 >0$ is a constant. Introduce a   filter  $\varsigma_4$ as follows
\begin{equation}
\begin{split}
\tau \dot{\varsigma_4}   + \varsigma_4 =\ &  \nu_4; \ \ \text{where} \ \ \varsigma_4 (0) =  \nu_4 (0)
\end{split}
\end{equation}
where $\tau$ is a time constant. Define an error $e_4 = \varsigma_4  - \nu_4$.  Now, choose $\xi_8  $ as follows:
\begin{equation}
\xi_8   = x_8- \varsigma_4  - x_8^c \label{eq:z_2}
\end{equation}
Introduce  (\ref{eq:z_2}) into (\ref{eq:z_1_dot})  and simplify
\begin{equation}
\dot{\xi}_7 = -p_4 \xi_7  + \xi_8   + e_4 + x_8^c   \label{eq:z_1_dot_modified}
\end{equation}
Take derivative of (\ref{eq:z_2})  
\begin{equation}
\dot{\xi}_8 = U_x - \dot{x}_8^c - {\nu_4 }/{\tau} + {\varsigma_4}/{\tau} \label{eq:z_2_dot}
\end{equation}
Choose a Lyapunov candidate function   as follows 
\begin{equation}
V_{\text{P}} = 0.5 ( \xi_7^2 + \xi_8^2 +  e_4^2 ) \label{eq:Lyap_fn}
\end{equation} 
 where $\alpha_x$ is a constant.   Next,   derivative of (\ref{eq:Lyap_fn}) yields
\begin{equation}
\dot{V}  = \xi_7 \dot{\xi}_7 + \xi_8\dot{\xi_8}_ + e_4 \dot{e}_4 
\end{equation}
Now, substituting (\ref{eq:z_1_dot_modified}), (\ref{eq:z_2_dot}) and using  $e_4 = \varsigma_4- \nu_4$ yields
\begin{equation}
\begin{split}
\dot{V}_{\text{P}} 
= \ & -p_4 \xi_7^2 + \xi_7 \xi_8 + \xi_7 e_4 + \xi_7 x_8^c  + \xi_8 \big(U_x- \dot{x}_8^c    - ( \nu_4 + \varsigma_4  )/{\tau} \big) - {e_4^2}/{\tau} - F_{\text{P}}(\star) e_4 
\end{split} \label{eq:derivative_of_LF}
\end{equation}
where $  F_{\text{P}}(x_7,x_8,\xi_7,\xi_8,e_4) = \frac{\partial \nu_4 }{\partial x_7} \dot{x}_7  + \frac{\partial \nu_4}{\partial \xi_7} \dot{\xi}_7$ is a continuous function bounded by $F_{\text{P}}(\star)$ . 
\begin{equation}
U_x = -    \xi_7  + \dot{x}_8^c + \left( {\nu}_4 - \varsigma_4 \right)/{\tau}  - k_4 \xi_8    \label{eq:standard_BSC}
\end{equation}
Substitute (\ref{eq:standard_BSC}) into (\ref{eq:derivative_of_LF}) and use Young's inequality  
\begin{equation}
\begin{split}
\dot{V}_{\text{P}}
\leq   & - \left( p_4 -1/2 \right) \xi_7^2   - \left( k_4 -1/2\right)  \xi_8^2      -  \left(2/ {\tau} - 1/2 \right) e_4^2    + B_{\text{x}} /2
\end{split}
\end{equation}
where $B_{\text{x}} \geq \text{max} \{ 3 (x_8^c)^2     + \bar{F}_{\text{P}}^2   \}.$ Choosing $p_4 > 0$, $k_4 > 0$ and $\tau\in (0,1] $, makes $\dot{V}_{\text{P}}$ negative. Next, for state estimation, an HGO is designed as follows \citep{khalil2007high}:
\begin{equation}
\begin{split}
\dot{\hat{x}}_7 & = \hat{x}_8 + {\beta_{7}}\tilde{x}_7 /{\varepsilon}\\
\dot{\hat{x}}_8 &  =  \hat{f}_4 (\hat{x},U_p ) + {\beta_{ 8}} \tilde{x}_7/{\varepsilon^2}
\end{split}
\end{equation}
where $\tilde{x}_7= \left( y_x - \hat{x}_7 \right)$ and $y_x=x_7$. In addition,  $\beta_{7}>0$ and $\beta_{8}>0$ are HGO gains. Furthermore, $\hat{f}_4(\hat{x},U_p ) = \left( \cos \hat{x}_1 \sin \hat{x}_3 \cos \hat{x}_5 + \sin \hat{x}_1 \sin \hat{x}_3\right) \frac{U_p}{m} $.
 
\textbf{Remark 1:} Position control requires information of estimated attitude angles. \newline 
\textbf{Remark 2:} With similar control design method, $U_y$ and $U_z$ for quadrotor position along $y$-axis and $z$-axis, respectively, can be designed. Once the auxiliary controllers are obtained, the thrust input $U_p$ is computed as:

 \begin{equation} \label{eq:Pos_U1}
 \begin{split}
U_p &= \frac{m \left(U_z + g \right)}{\cos \phi_{\text{des}} \cos \theta_{\text{des}}} \\
\theta_{\text{des}} &=  {\tan}^{-1} \left( \frac{U_x \cos \psi_{\text{des}} + U_y \sin \psi_{\text{des}}}{U_z + g} \right)\\
  \phi_{\text{des}} & =  \tan^{-1} \left( \frac{ \left( U_x \sin \psi_{\text{des}} - U_y \cos \psi_{\text{des}}\right) \cos \theta_{\text{des}} }{U_z+g} \right)
  \end{split}
\end{equation}
where $\psi_{\text{des}}$ is  desired yaw angle.
\begin{theorem}
Desired trajectory tracking control performance for the quadrotor with dynamical model in (\ref{eq:Quadrotor_Dynamical_Model}) can be achieved by designing the attitude and position control, along with appropriate control gain design of the controller, DO, and HGO for tracking, disturbance attenuation, and state estimation, respectively.
\end{theorem}
\begin{proof}
The proof follows similar steps of theorem 1. Hence, it is omitted here.
\end{proof}

\section{Simulations}

This section presents simulations of the closed-loop control algorithm applied to the quadrotor DJI F450. To analyze the effectiveness and major contributions of the research results, these simulations are compared with the control algorithm presented by \citep{siddiqui2024model}. The comparison is shown in Figure \ref{fig:Traj_tracking}.

\subsection{Desired trajectory and unknown disturbances}
The aim is to achieve trajectory tracking control in the 3D XYZ plane. The desired trajectory is chosen as follows \citep{siddiqui2024model}:
\begin{equation}
r(t) = \begin{bmatrix}
x_r (t) & y_{r} (t) &z_{r} (t)
\end{bmatrix}^T = \begin{bmatrix}
3 - 3 \cos \left(\frac{t}{15} \right) & 2 + 3 \sin \left( \frac{t}{15} \right) & 1 + \frac{1}{10} t
\end{bmatrix}^T
\end{equation}
Unlike \cite{siddiqui2024model}, who considered only attitude disturbances, we assume disturbances in both position and attitude models of the quadrotor. Position disturbances include sinusoidal, constant, and chirp forms given by $d_x(t) = \sin(0.1t)$, $d_y(t) = 1$ at 50 seconds, $d_z(t) = 10^{-1} + 10^{-2}t$ up to 100 seconds. Attitude disturbances in roll, pitch, and yaw use Simulink blocks to obtain Gaussian, uniform, and band-limited white-noise, sampled every 15 s and 1 s, respectively.

\begin{table} 
  \centering
  \caption{Control parameters. \\ \textit{Source: Author’s own work.}  }
  \label{tab:control_params}
    \begin{tabular}{ l |  l  l  l  l  l    l    }
    \hline 

    Parameter / Model    & Roll & Pitch & Yaw  &  $x $& $y$ & $z$ \\ \hline
    Filter gains: $p_1$ & $100$ & $ 100$ & $1$  &  $0.1$&   $0.1$& $0.1$ \\  
    DSC gains:  $k_1$& $120$ & $ 120$ & $10$  &  $5$&  $5$& $1$ \\  
    HGO gains: $\beta_1$& $1$ & $ 1$ & $1$  &  $1$&  $1$& $1$ \\  
    HGO gains: $\beta_2$& $2$ & $ 2$ & $2$  &  $2$&  $2$& $2$ \\
    Command-filter gains: $m_1$& $1$ & $ 1$ & $1$  &  $1$&  $1$& $1$ \\
    Command-filter gains: $m_2$& $1$ & $ 1$ & $1$  &  $0.1$&  $0.1$& $0.1$ \\  \hline

    \end{tabular}
\end{table}

\subsection{Simulation results and discussion}
\subsubsection{Trajectory tracking and quadrotor outputs}

\begin{figure}[H]
  \centering
    \includegraphics[width=1\linewidth ,clip,keepaspectratio]{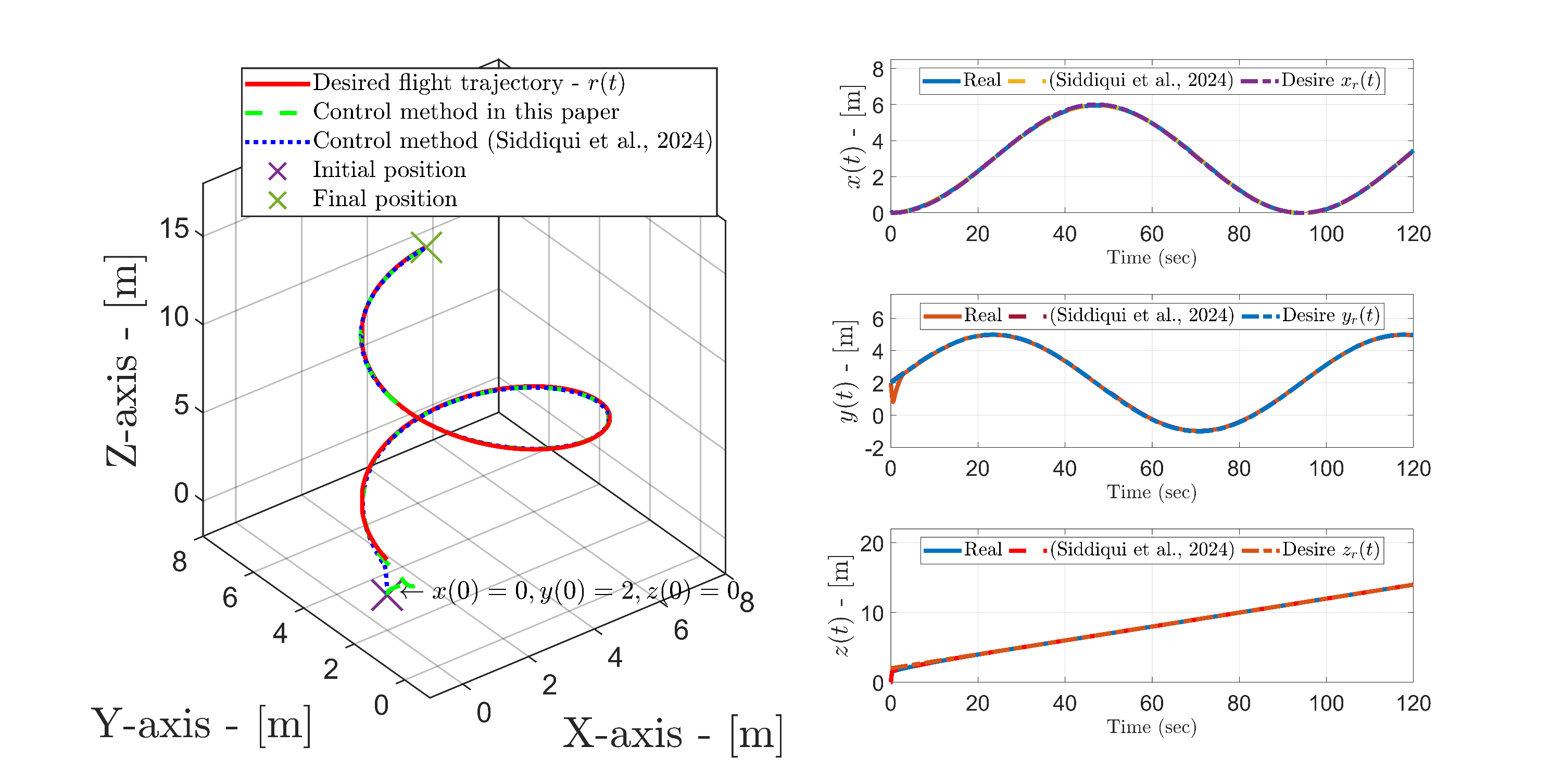}
    \caption{Trajectory tracking in the Cartesian plane and quadrotor position.\\ \textit{Source: Author’s own work; elements adapted from \citep{siddiqui2024model}}}
    \label{fig:Traj_tracking}
\end{figure}

Figure \ref{fig:Traj_tracking} illustrates the successful trajectory tracking of the quadrotor flight mission over $2$ minutes. The proposed control technique in this study shows no overshoot during takeoff, whereas the compared paper exhibits a slight overshoot. The comparison of the quadrotor position along each axis in the Cartesian plane is also shown. It shows strong trajectory tracking under disturbances. Table \ref{tab:control_params} lists the control gains designed for the control system.

\begin{figure}[H]
  \centering
    \includegraphics[width=1\linewidth ,clip,keepaspectratio]{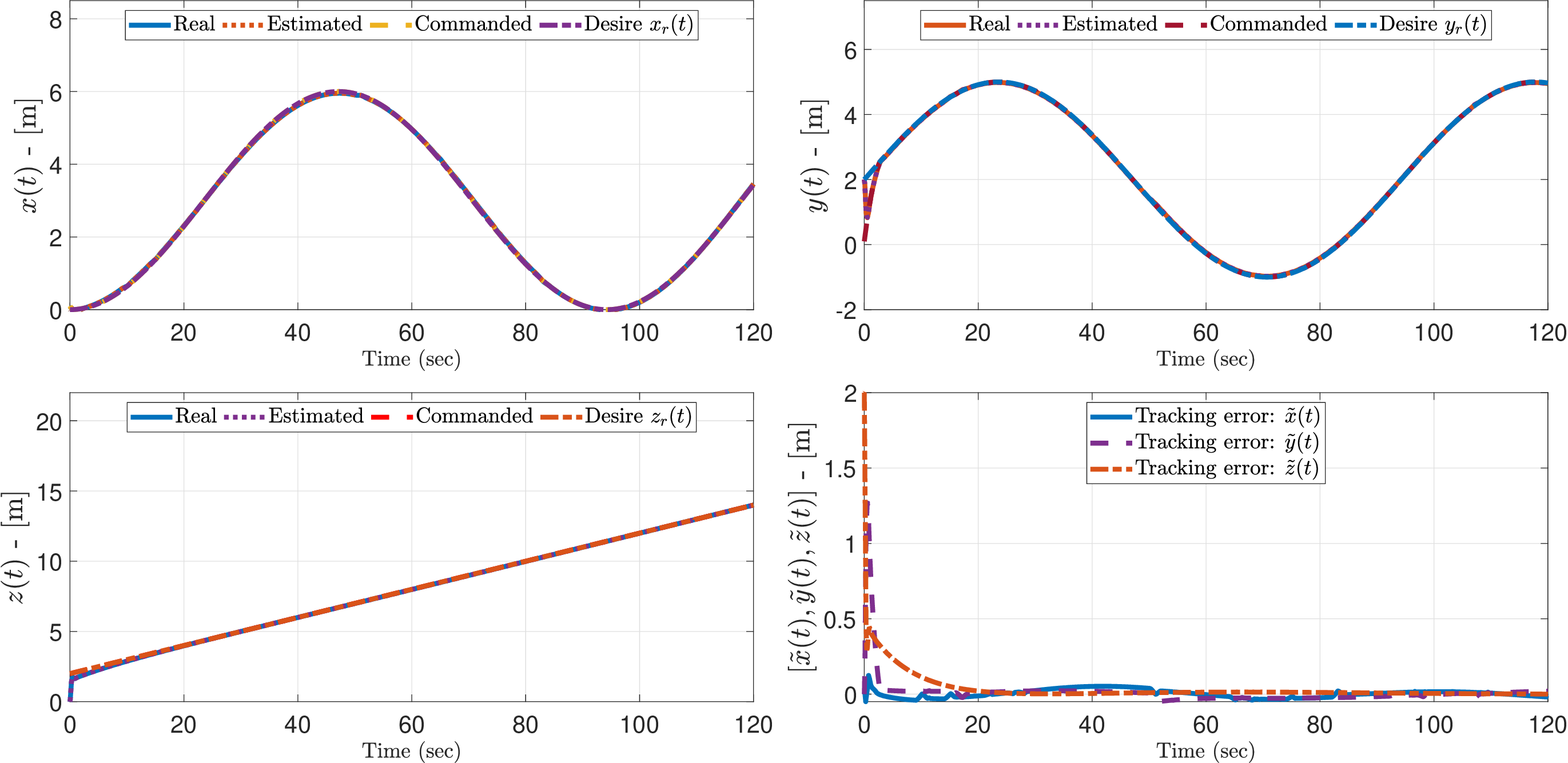}
    \caption{Quadrotor position and tracking errors.\\ \textit{Source: Author’s own work.} }
    \label{fig:Position_and_Error}
\end{figure}

\begin{figure}[H]
  \centering
    \includegraphics[width=1\linewidth ,clip,keepaspectratio]{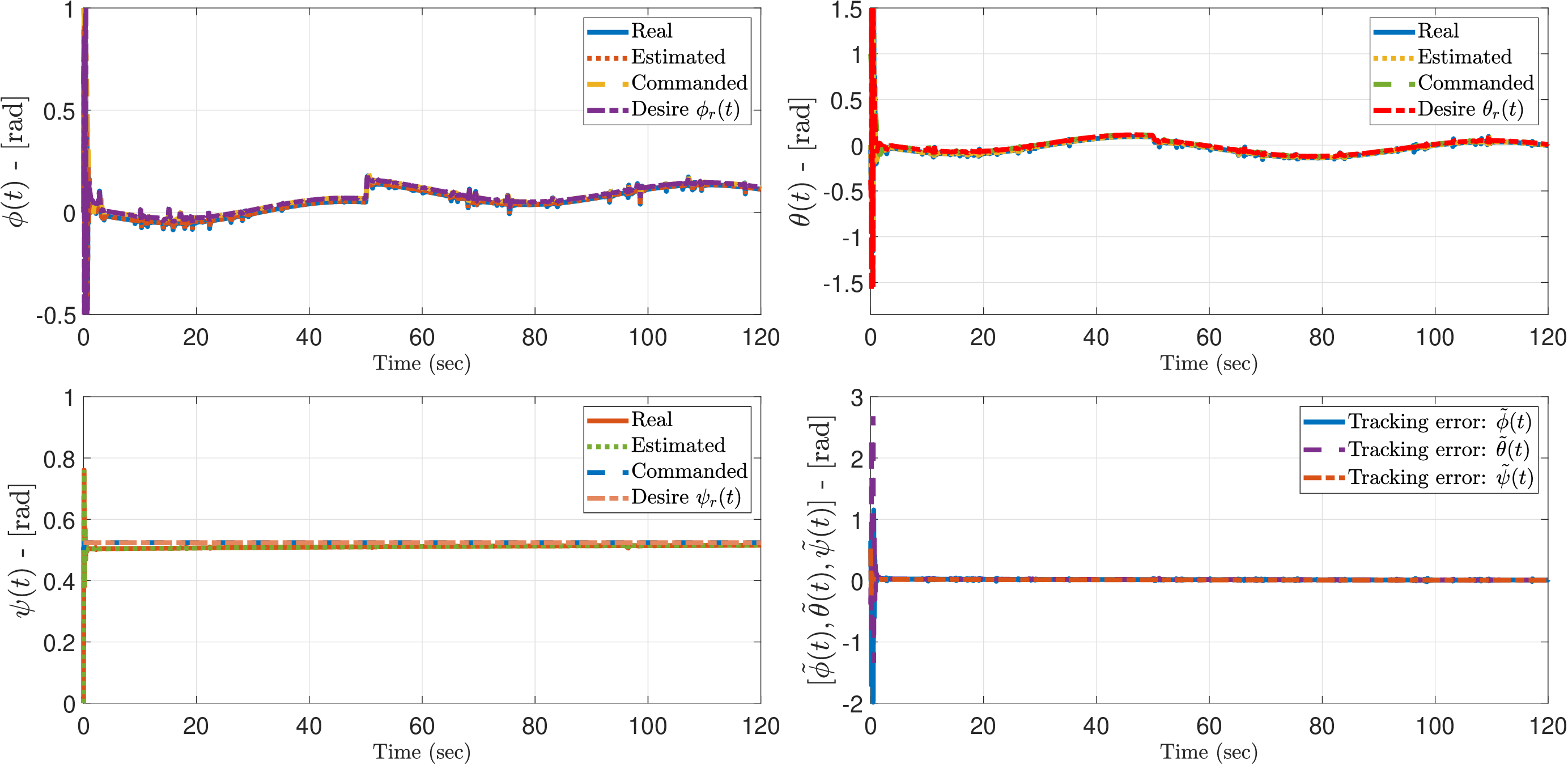}
    \caption{Quadrotor attitude  and tracking error. \\ \textit{Source: Author’s own work.}}
    \label{fig:Attitude_and_error}
\end{figure}

 Detailed comparisons of each quadrotor position and attitude are shown in Figure \ref{fig:Position_and_Error}
 and Figure \ref{fig:Attitude_and_error}, respectively. We observe that the quadrotor achieves vertical takeoff at a continuous rate while following a spiral trajectory. Furthermore, it is evident that the tracking errors are reduced and stay within a neighborhood of the origin throughout the flight mission in the presence of disturbances.


\begin{figure}[H]
  \centering
    \includegraphics[width=0.45\linewidth ,clip,keepaspectratio]{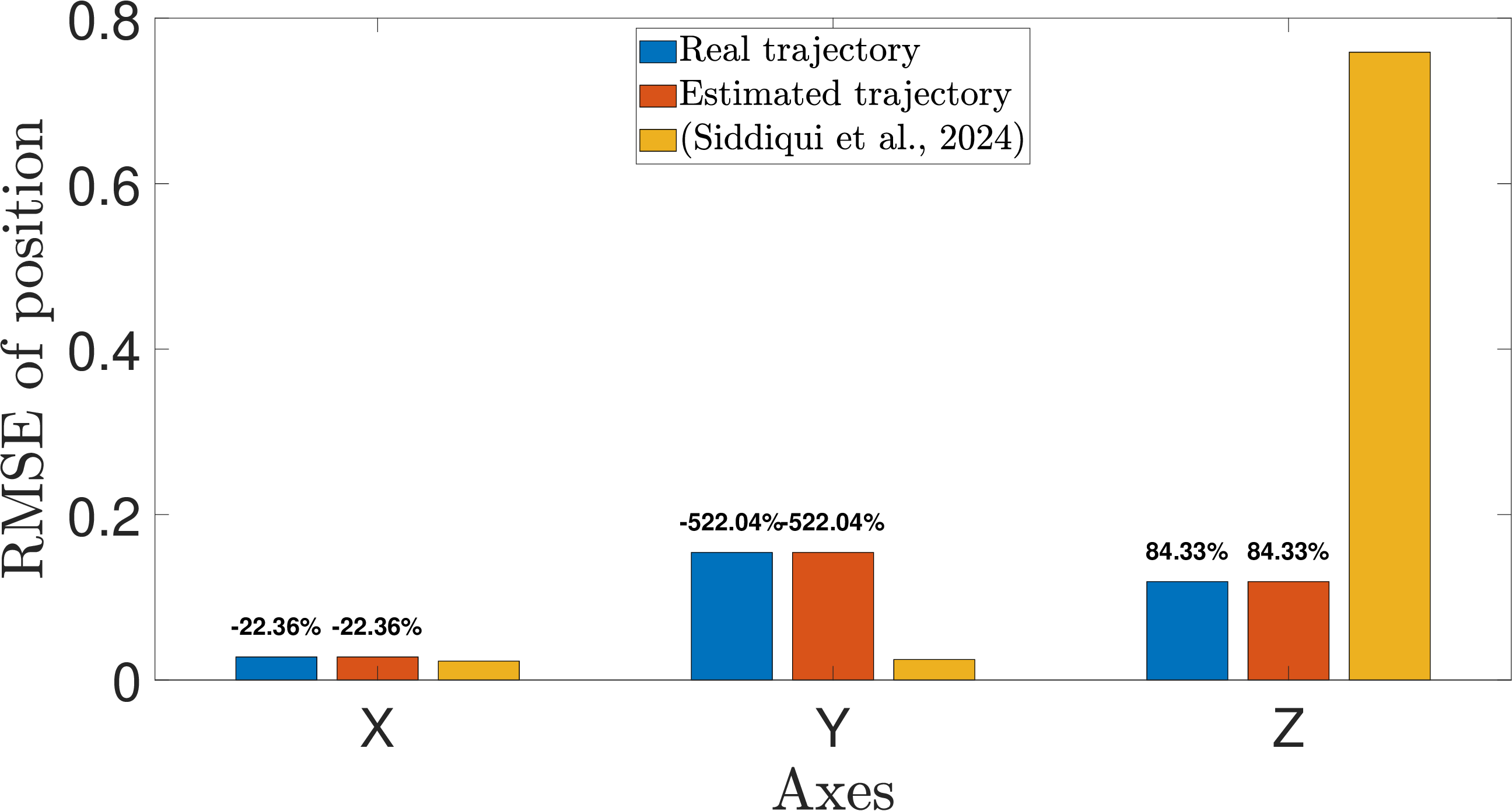}
    \includegraphics[width=0.45\linewidth ,clip,keepaspectratio]{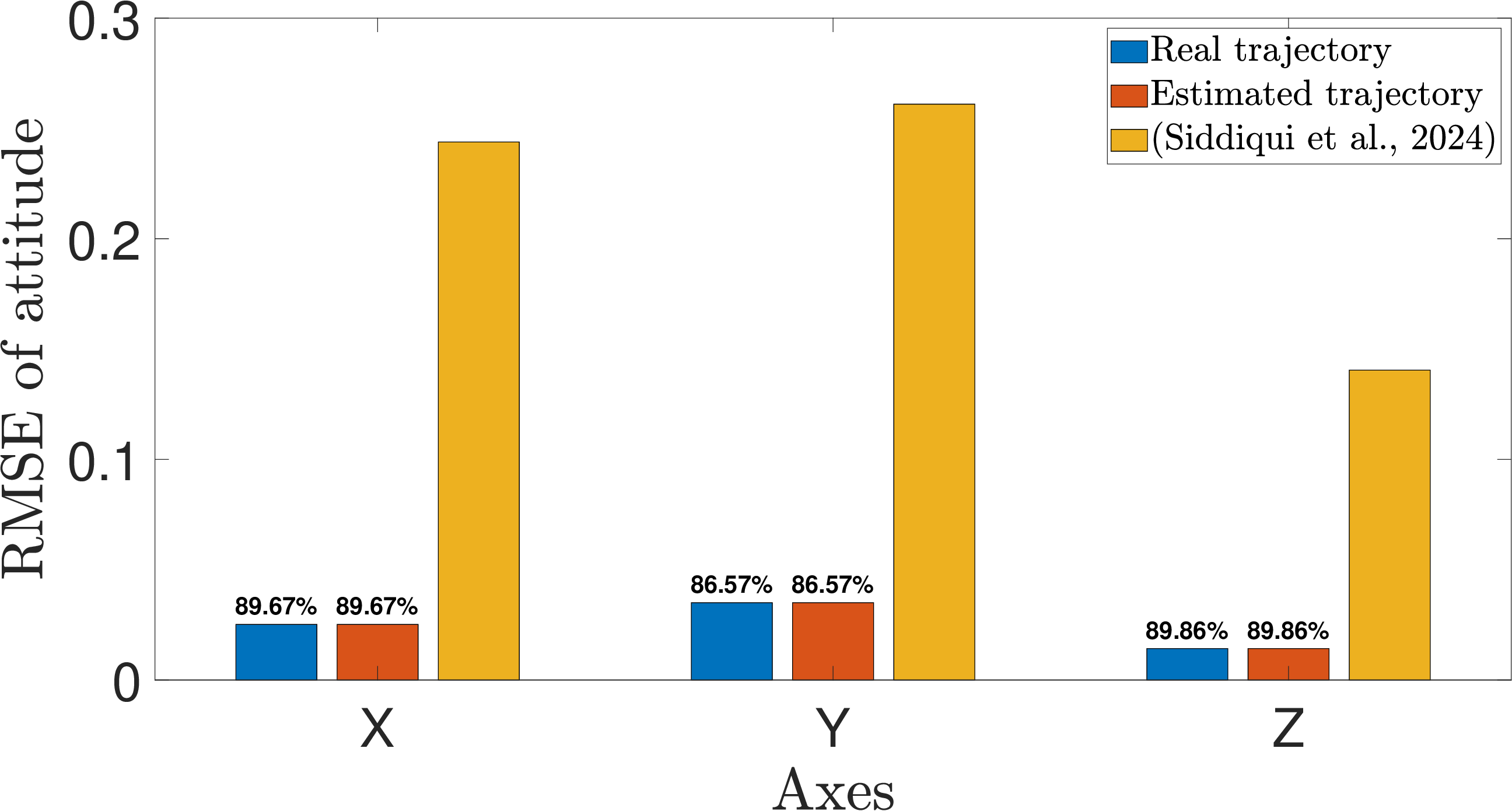}
    \caption{Root mean square tracking errors of position and attitude. \\ \textit{Source: Author’s own work.} }
   \label{fig:BAR_charts}
\end{figure}

Figure~\ref{fig:BAR_charts} presents bar charts depicting the root mean square errors (RMSE) of position and attitude for the quadrotor, in comparison with the results reported in \citep{siddiqui2024model}. We observe a substantial reduction in both the real and estimated attitude RMSEs of the quadrotor. Regarding position, a significant decrease in RMSE, approximately $90\%$, is observed along the z-axis (altitude) for both the real and estimated positions. In contrast, no notable reduction is observed in the RMSEs along the x-axis and y-axis. However, the tracking errors along these axes remain minimal and confined within a small neighborhood around the origin, thereby ensuring effective trajectory tracking.

\begin{figure}[H]
  \centering
    \includegraphics[width=1\linewidth ,clip,keepaspectratio]{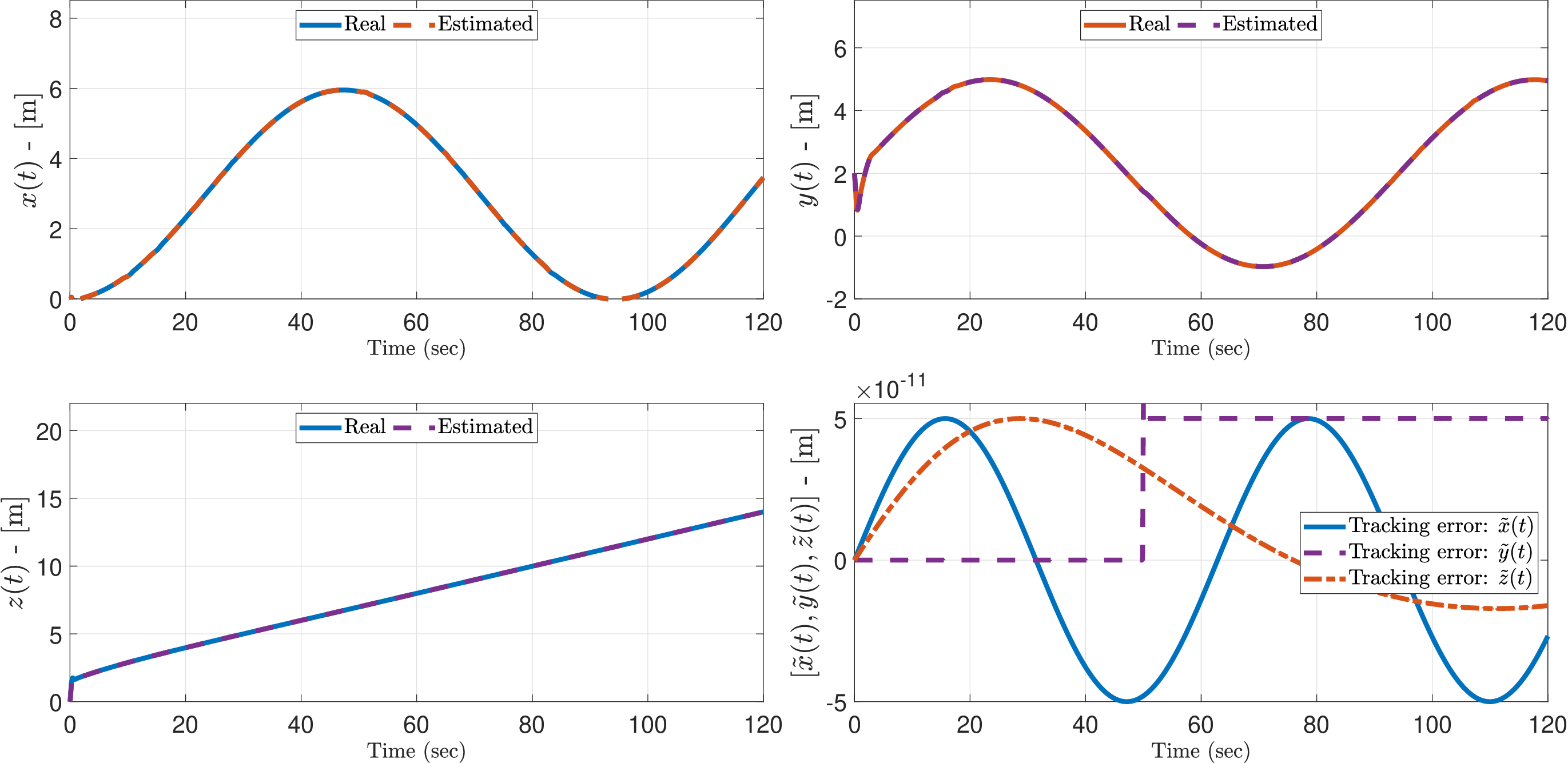}
    \caption{Quadrotor estimated position and estimation errors.\\ \textit{Source: Author’s own work.} }
    \label{fig:State_est_Position_and_Error}
\end{figure}

\begin{figure}[H]
  \centering
    \includegraphics[width=1\linewidth ,clip,keepaspectratio]{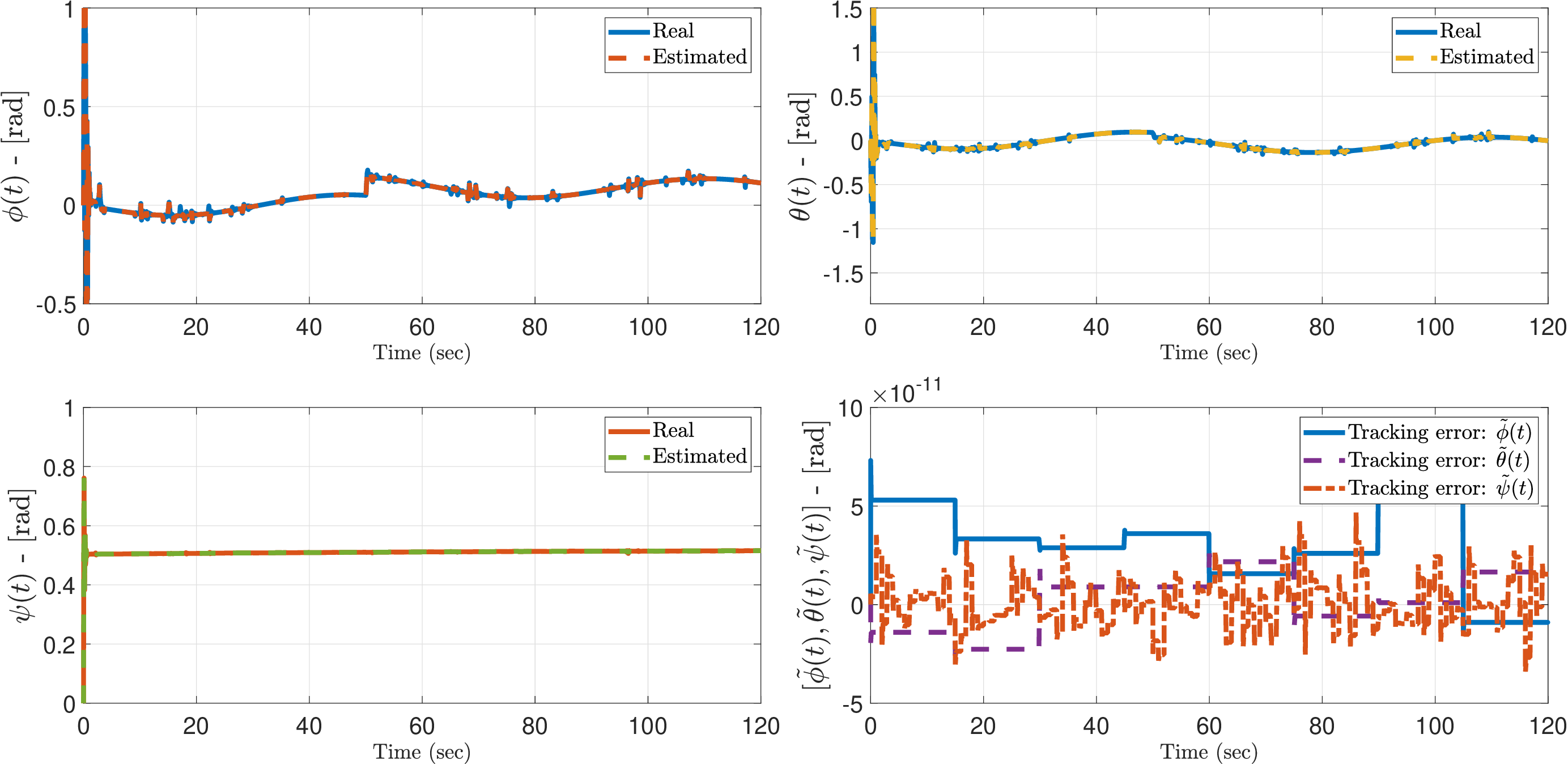}
    \caption{Quadrotor estimated attitude and estimation errors. \\ \textit{Source: Author’s own work.} }
    \label{fig:State_est_Attitude_and_error}
\end{figure}

\subsubsection{State and DO}
The state-estimation simulation results are presented in Figure~\ref{fig:State_est_Position_and_Error} and Figure~\ref{fig:State_est_Attitude_and_error}, corresponding to the quadrotor's position and attitude, respectively. These results demonstrate that the real quadrotor outputs are accurately estimated using the HGO  state-estimation method. Additionally, the estimation errors are shown, confirming that the HGO-based estimation technique recovers the real state variables, even in the presence of disturbances encountered during the quadrotor’s flight mission.

\begin{figure}[H]
  \centering
    \includegraphics[width=1\linewidth, clip,keepaspectratio]{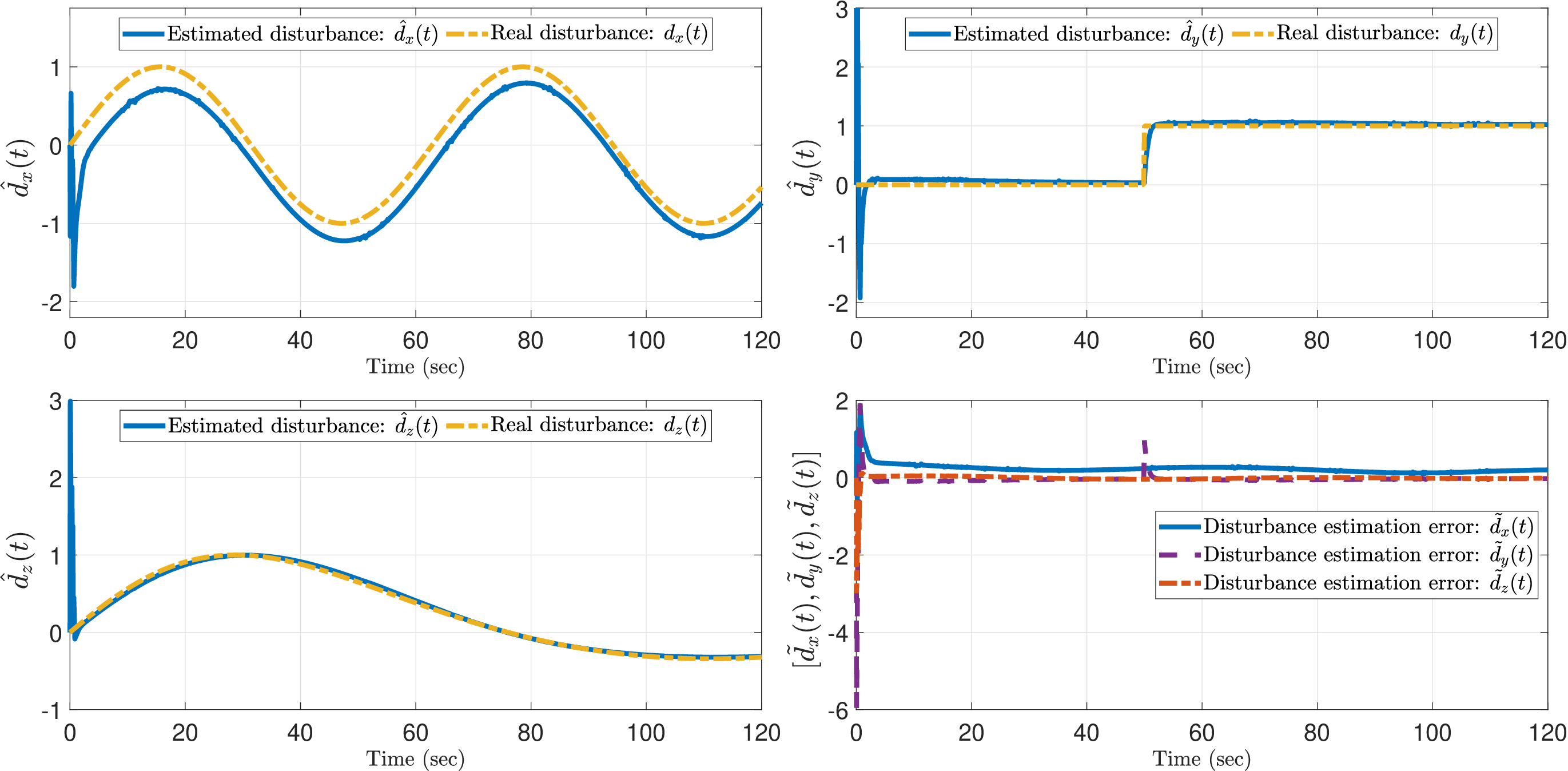}
    \caption{Disturbance estimation in quadrotor position. \\ \textit{Source: Author’s own work.}}
    \label{fig:DO_Position}
\end{figure}

\begin{figure}[H]
  \centering
    \includegraphics[width=1\linewidth, clip,keepaspectratio]{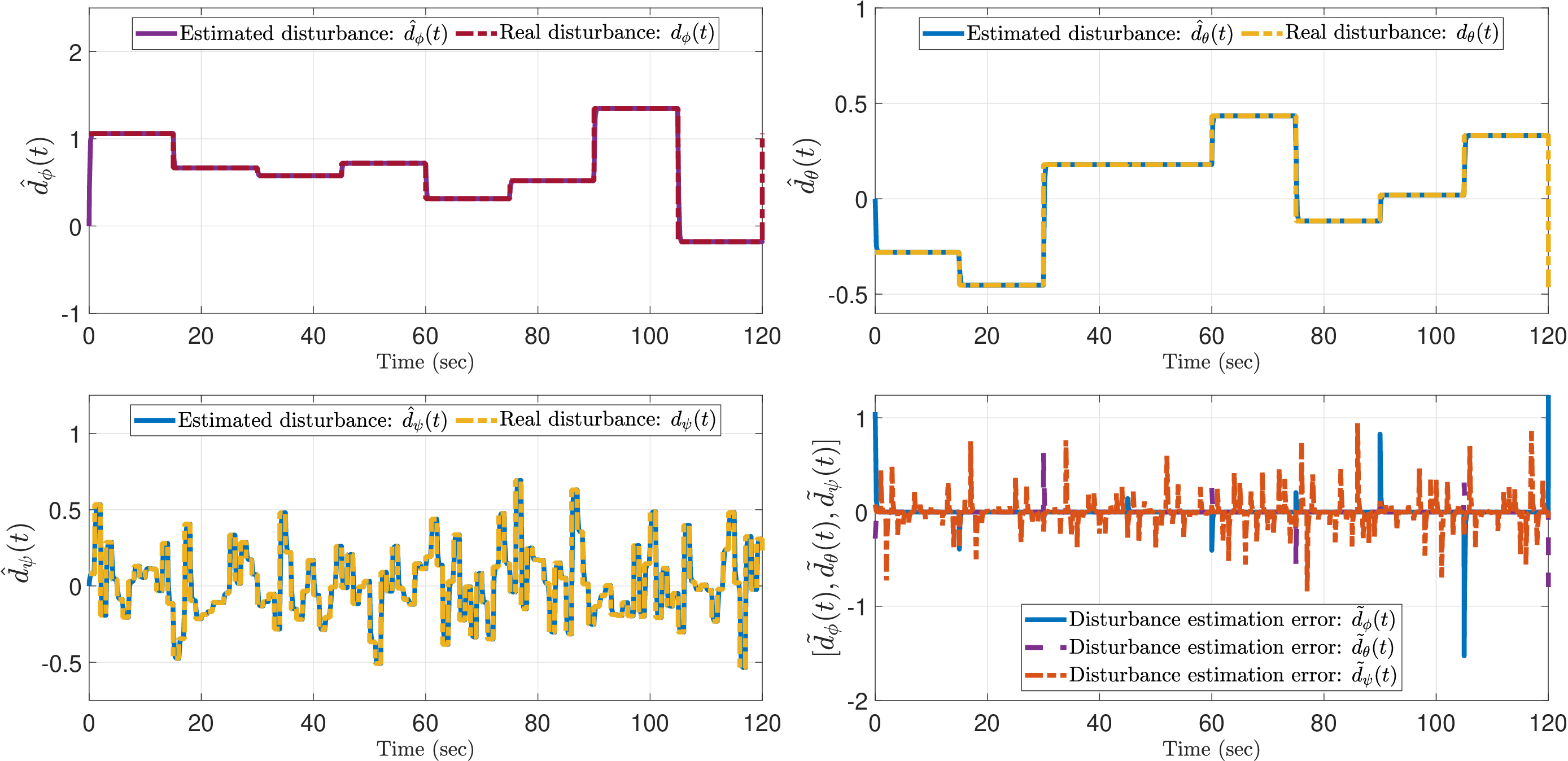}
    \caption{Disturbance estimation in quadrotor attitude. \\ \textit{Source: Author’s own work.}}
    \label{fig:DO_Attitude}
\end{figure}

Figure~\ref{fig:DO_Position} and Figure~\ref{fig:DO_Attitude} depict the disturbance estimation in the position and attitude subsystems of the quadrotor model. External disturbances are assumed to act on the position subsystem, while the attitude subsystem is subject to internal disturbances. The observer estimates these disturbances accurately. Once the control algorithm is finalized, the estimated disturbances are incorporated into the control laws to implement the DOBC strategy. The disturbance sampling time is adjustable; here we use a longer interval to improve plot readability without inducing chattering, most noticeable in the yaw channel. The disturbance estimation errors, also depicted in the figures, indicate that the errors converge asymptotically to zero in the position model. In the case of the attitude DO, the observed spikes are attributed to sudden changes in the disturbances, which are nonetheless accurately estimated.

\subsubsection{Control inputs}
\begin{figure}[H]
  \centering
    \includegraphics[width=1\linewidth ,clip,keepaspectratio]{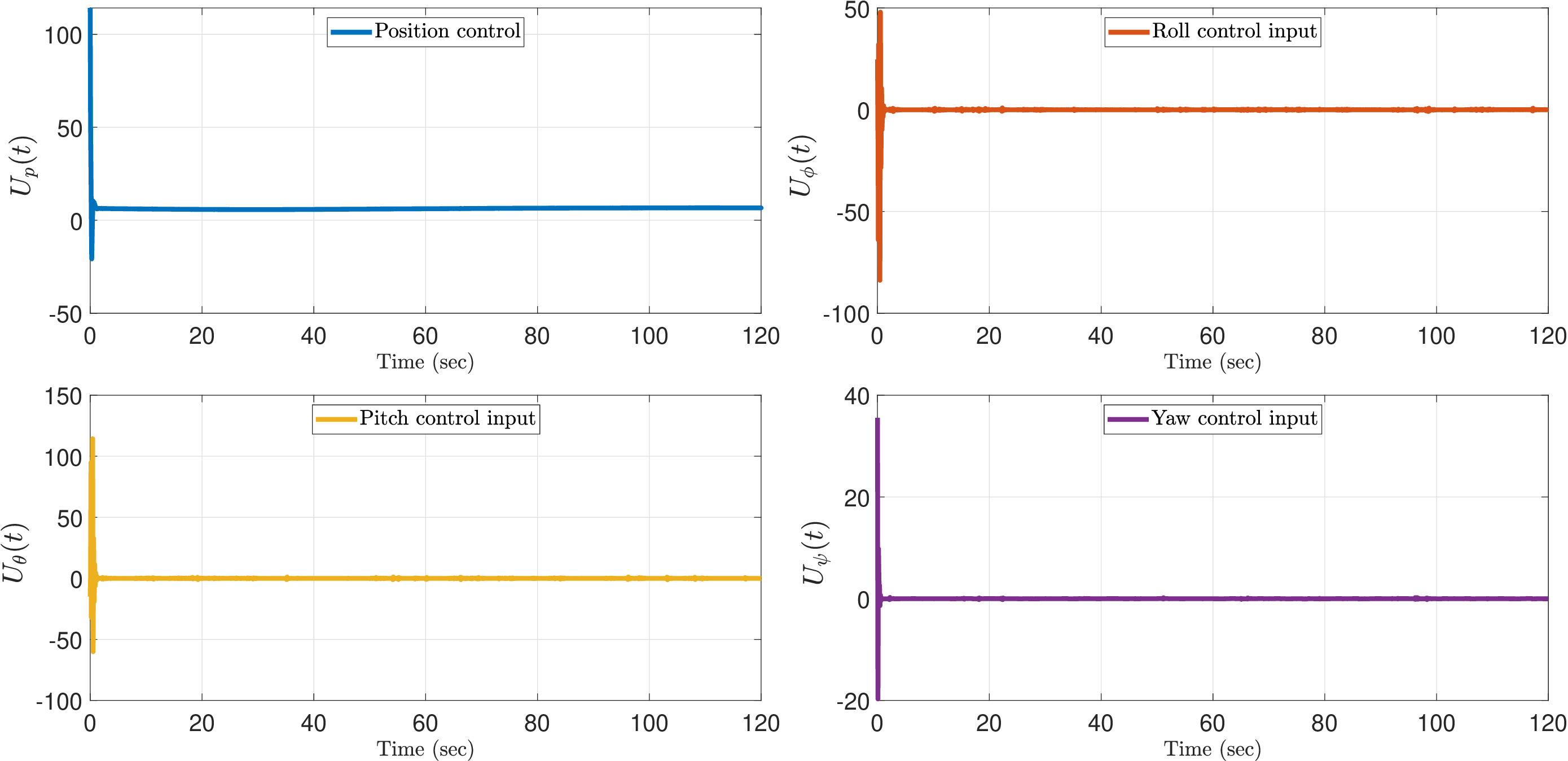}
    \caption{Control inputs. \\ \textit{Source: Author’s own work.}}
    \label{fig:Control_inputs}
\end{figure}

\begin{figure}[H]
  \centering
    \includegraphics[width=1\linewidth ,clip,keepaspectratio]{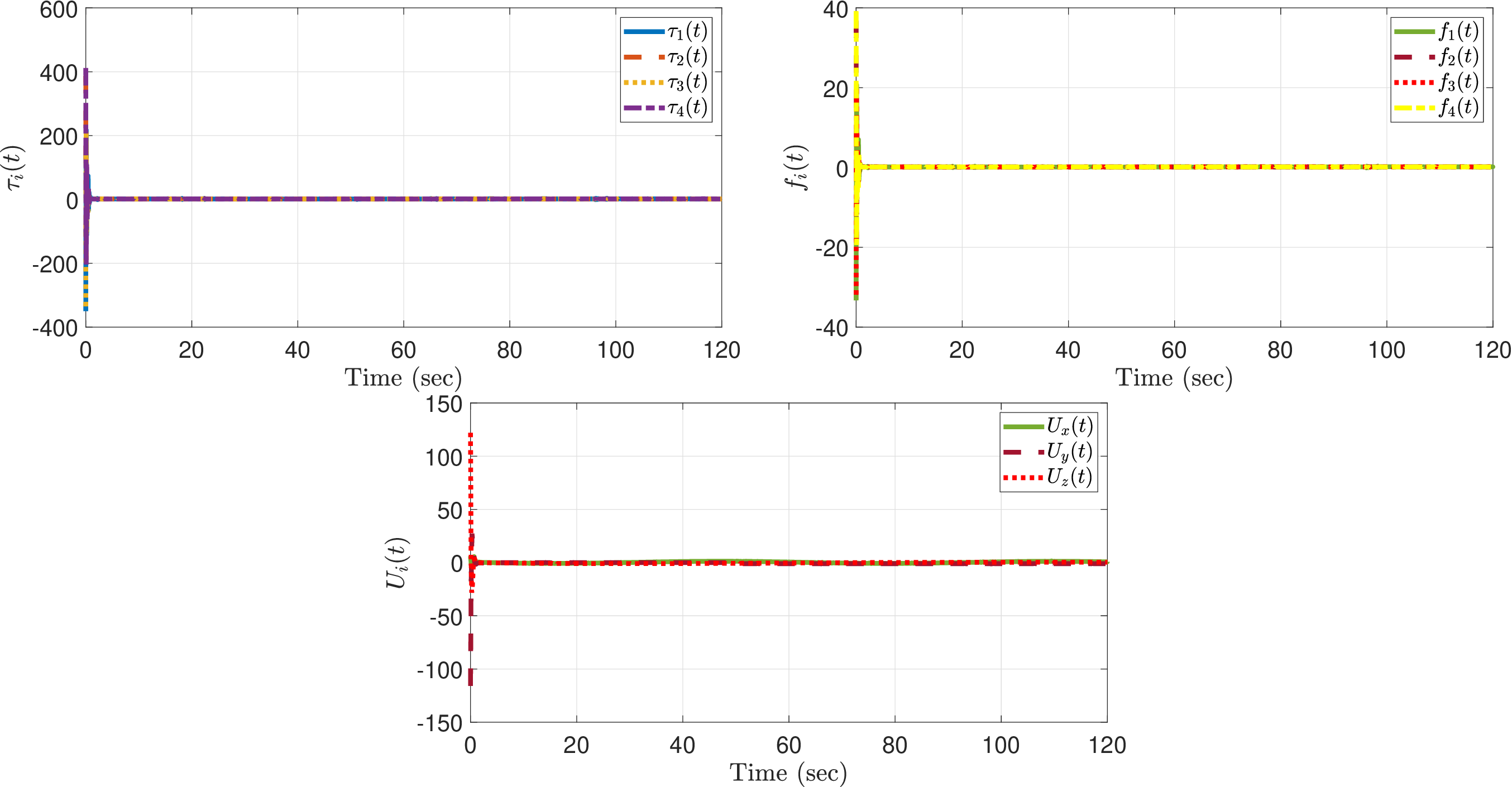}
    \caption{Torque and forces of each rotor, virtual position control inputs. \\ \textit{Source: Author’s own work.} }
    \label{fig:Force_torque}
\end{figure}

The quadrotor is underactuated, relying on four inputs, roll, pitch, yaw, and thrust, for full maneuvering. Position control is achieved through the use of three virtual control inputs. Following trajectory tracking simulations, the control inputs required to accomplish the desired flight mission are shown in Figure~\ref{fig:Control_inputs}. Furthermore, Figure~\ref{fig:Force_torque} presents the torque and thrust generated by each rotor, computed using $\tau_i = b \omega_i^2$ and $f_i = b \omega_i^2$, respectively, where $i \in \{1,2,3,4\}$ and $\omega_i$ is determined from Equation~(\ref{eq:Control_inputs}). Additionally, since the position subsystem is underactuated with a single control input, the corresponding virtual control inputs used to achieve position regulation are shown in the figure.

\subsection{Discussion on research contributions and limitations}

The method is suitable for real-time use: it requires modest processing power and still maintains stability when the model or external forces vary. Unknown disturbances acting on both position and attitude channels are estimated and rejected, and reliance on high-precision sensors is reduced by the HGO. Simulated flights report marked reductions in altitude RMSE and noticeable improvements along the horizontal axes. The filter lowers computational load and limits actuator chattering. A remaining drawback is the transient peaking caused by the observer; future work should explore softer gain schedules.

The filter also keeps the backstepping math simple by replacing repeated derivatives of states and virtual controls. The incorporation of the command filter facilitates numerical differentiation of the desired state variables, mitigating delays associated with direct differentiation methods.
 Additionally, the developed controller minimizes chattering, thereby reducing heat losses and wear on actuators.

 Although extensive research has been dedicated to refining this control method, it relies on HGO state-estimation criteria which may introduce finite-time escape and instability concerns due to the reciprocal gain parameter $\varepsilon$. \citep{siddiqui2024model} addressed this issue by designing low-power observer-based state estimation criteria, but it is suitable for integral chain systems.  Thus, the presented research work can be extended to address this issue. Nevertheless, the presented control method achieves strong results for the trajectory tracking of a quadrotor with unknown disturbances. Thus, it can be implemented on hardware in future work, similar to \cite{zhao2025command}.
 
Moreover, the filtering scheme used in this research combines a command filter with a first-order filter. The command filter provides the required derivatives without direct differentiation, whereas the first-order filter prevents repeated differentiation of auxiliary inputs in the backstepping design, thereby avoiding the well-known \textit{explosion of complexity}. These techniques are effective for quadrotor control and are also applicable to systems in which velocity measurements are unavailable or unreliable and where direct differentiation is computationally expensive. Hence, the proposed approach is generic and can be extended to other dynamic systems.
 
\section{Conclusion} 
We present a robust trajectory-tracking control strategy for underactuated quadrotor systems, designed to accomplish nonlinear flight missions in the Cartesian plane. 
The proposed controller integrates a nonlinear DO to estimate multiple unknown disturbances and uses a high-gain state observer to reduce dependence on high-fidelity sensor measurements. 
A backstepping approach is employed for controller design, where the \textit{explosion of complexity} problem is addressed through the use of a first-order filter and a command filter for numerical differentiation of the desired trajectory. 
Simulation results indicate that the proposed method achieves superior tracking accuracy under uncertain conditions, with approximately 90\% reduction in $z$-axis RMSE and improvements in attitude regulation compared to existing approaches. 
Closed-loop stability is established through Lyapunov-based analysis, establishing rigorous mathematical validation of the control laws. 

Despite these advantages, some limitations remain. 
The method assumes accurate modeling of motor dynamics and relies on numerical filtering, which may introduce computational overhead that affects real-time performance. 
In addition, while the disturbance observer estimates nonlinear and stochastic disturbances, its performance may degrade under noisy sensor conditions under high-frequency sensor noise. 

Future research will focus on addressing these limitations by improving numerical differentiation techniques, incorporating reduced-order or adaptive observers to further enhance robustness, and extending the proposed strategy to experimental validation on hardware platforms. Such efforts will broaden the applicability of the proposed control framework for real-time quadrotor missions in uncertain and dynamic environments.

\nocite{*}
\bibliographystyle{agsm}
\bibliography{ieee_references}

\end{document}